# High-throughput validation of ceRNA regulatory networks

Hua-Sheng Chiu[1], María Rodríguez Martínez[2], Mukesh Bansal[3], Aravind Subramanian[4], Todd R. Golub[4,5], Xuerui Yang[6#], Pavel Sumazin[1#] & Andrea Califano[3,7#]

[1] Texas Children's Cancer Center and Department of Pediatrics, Baylor College of Medicine, Houston, Texas, USA

[2] IBM Research—Zurich, 8803 Rüschlikon, Zurich, Switzerland

[3] Columbia Department of Systems Biology, Center for Computational Biology and Bioinformatics, Herbert Irving Comprehensive Cancer Center, Columbia University, New York, New York, 10032, USA

[4] Broad Institute, 7 Cambridge Center, Cambridge MA, 02142, USA

[5] Dana-Farber Cancer Institute, Boston, MA 02115, USA; and Howard Hughes Medical Institute, Chevy Chase, MD 20815-6789, USA

[6] MOE Key Laboratory of Bioinformatics, Tsinghua-Peking Center for Life Sciences, School of Life Sciences, Tsinghua University, Beijing 100084, China

[7] Department of Biomedical Informatics, and Department of Biochemistry and Molecular Biophysics, and Institute for Cancer Genetics, Columbia University, Herbert Irving Comprehensive Cancer Center, Columbia University, New York, New York, 10032, USA

Correspondence: yangxuerui@tsinghua.edu.cn, sumazin@bcm.edu, califano@c2b2.columbia.edu

# Abstract

**Background**: MicroRNAs (miRNAs) play multiple roles in tumor biology [1]. Interestingly, reports from multiple groups suggest that miRNA targets may be coupled through competitive stoichiometric sequestration [2]. Specifically, computational models predicted [3, 4] and experimental assays confirmed [5, 6] that miRNA activity is dependent on miRNA target abundance, and consequently, changes to the abundance of some miRNA targets lead to changes to the regulation and abundance of their other targets. The resulting indirect regulatory influence between miRNA targets resembles competition and has been dubbed competitive endogenous RNA (ceRNA) [5, 7, 8]. Recent studies have questioned the physiological relevance of ceRNA interactions [9], researchers ability to accurately predict these interactions [10], and the number of genes that are impacted by ceRNA interactions in specific cellular contexts [11].

**Results**: To address these concerns, we reverse engineered ceRNA networks (ceRNETs) in breast and prostate adenocarcinomas using context-specific TCGA profiles [12-14], and tested whether ceRNA interactions can predict the effects of RNAi-mediated gene silencing perturbations in PC3 and MCF7 cells. Our results, based on tests of thousands of inferred ceRNA interactions that are predicted to alter hundreds of cancer genes in each of the two tumor contexts, confirmed statistically significant effects for half of the predicted targets.

**Conclusions**: Our results suggest that the expression of a significant fraction of cancer genes may be regulated by ceRNA interactions in each of the two tumor contexts.

# Background

Identifying regulatory interactions that mediate the effects of genomic alterations is a necessary step for interpreting the function of trans-acting variants in complex diseases, including cancer [15, 16]. Among these, miRNA dysregulation, arising from alterations targeting their transcriptional [17] or biogenesis regulators [18], plays an established role in tumorigenesis [1]. Recently, multiple groups have reported on gene products that modulate miRNA activity [5-7, 19-26], including RNA species that can alter the abundance of other RNAs *in trans* through ceRNA interactions. These studies show that targets of the same miRNAs are coupled and that up- or down-regulation of one target may alter the expression of other cognate targets by sequestering or releasing their shared miRNA molecules, respectively (Figure 1A).

Since the discovery of ceRNA regulation in human cells [21, 22] multiple reports questioned the physiological relevance of ceRNA interactions, researcher's ability to predict them, and the number of genes that are affected in each context [9-11]. To address these concerns, we proceeded to test genome-wide ceRNA predictions made by the information-theoretic Hermes algorithm [5]. For the sake of generality, we performed this analysis in two distinct tumor contexts, using a set of large-scale and high-throughput shRNA-mediated perturbation assays in model cell lines assembled by the Library of Integrated Network-based Cellular Signatures (LINCS) [14]. Specifically, we inferred ceRNETs using TCGA profiles of prostate and breast adenocarcinomas and tested them using a LINCS compendium of perturbation profiles, representative of the shRNA-mediated silencing of >3,000 genes in PC3 and MCF7 cells [12-14]. We propose that the high validation rates of these assays can inform on the accuracy of computational predictions, and will help estimate the number of genes that are modulated by ceRNA in representative tumor contexts.

## Methods and Results

### Inference method

We used an extended version of the Hermes algorithm, which we had previously introduced to discover and validate glioma-specific ceRNAs [5], to systematically discover ceRNA interactions in prostate (PRAD) and breast (BRCA) adenocarcinomas, using matched miRNA and mRNA expression profiles of the corresponding TCGA cohorts. While the ceRNA inference component of the algorithm was unchanged, the new algorithm also supports the identification of the specific miRNAs that mediate each interaction (*mediators*); these miRNAs are predicted to target both mRNAs in a ceRNA interaction, and their activity is affected by modulation of target mRNA abundance. This extension is useful for generating more specific hypotheses for future functional testing.

We note that Hermes-inferred ceRNA interactions are independent of the co-expression of coupled mRNAs; rather, they are based on assessing whether the abundance of one mRNA species modulates abundance of the other (and vice-versa), via their shared miRNA program. This assessment is based on the statistical significance of the mutual information between the abundance of one mRNA species and one or more miRNA, given the abundance of another mRNA targeted by the same miRNAs. We note that a majority of predicted ceRNA interactions involved mRNAs that are not significantly co-expressed, and co-expression did not implicate genes as ceRNA interacting partners.

Hermes predicts ceRNA-coupling between two mRNAs based on the relative size of their shared miRNA regulatory program, as predicted by the Cupid algorithm [5], and the conditional mutual information between one of the mRNAs and each of their shared miRNAs, given the other mRNA. Namely, given genes $T_i$ and $T_j$, and the set of miRNAs that regulate them $\Pi_{miR}(T_i)$ and $\Pi_{miR}(T_j)$, their shared program is identified by taking the intersection $\Pi_{miR}(T_i; T_j) = \Pi_{miR}(T_i) \cap \Pi_{miR}(T_j)$. First, Hermes tests that the size of $\Pi_{miR}(T_i; T_j)$ relative to the sizes of the individual programs is statistically significant at FDR < 0.01 by Fisher's exact test (FET). Then, Hermes evaluates the statistical significance $p_{kij}$ (*p*-value) of the test $CMI[miR_k; T_i | T_j] > MI[miR_k; T_i]$, where CMI and MI stand for conditional mutual information and mutual information respectively, and the variables indicate the expression of the corresponding RNA species [5].

The CMI is estimated using an adaptive partitioning algorithm [27] by first iteratively partitioning the 3-dimentional expression space evenly into 8 partitions per iteration until partitions are balanced (p>0.05 by Chi-squared test), and then summing up CMI across partitions. *P*-values for each triplet are computed based on a null-hypothesis where the candidate modulator's expression ($T_j$) is shuffled 1,000 times, thus preserving the pairwise mutual information between miRNA and target. Final significance across the entire set of miRNA mediators is computed using Fisher's method to integrate both regulatory directions, i.e. $T_i$ affecting *miR$_k$* regulation of $T_j$ as well as $T_j$ affecting *miR$_k$* regulation of $T_j$, for all miRNA mediators $\Pi_{miR}(T_i; T_j)$. Specifically, $X^2 = -2\sum_{k=1}^{N}\ln(p_{kij}, p_{kji})$ is computed and used to estimate a significance p-value for the entire program. Note that $X^2$ follows a Chi-square distribution, with $4N$ degrees of freedom, where $N$ is the number of miRNAs in the shared program. Finally, only predictions passing significance of FDR<1E-3 are selected. Note that selected predictions by Hermes have been previously validated in glioblastoma cell lines [5]. In addition, the presence of transcripts with alternative 3' UTRs is expected to reduce the sensitivity of prediction.

In order to identify miRNA mediators in addition to ceRNA interactions, we modified Hermes to perform greedy addition of miRNA mediators and to optimize the combined *p*-value for each predicted interaction. Namely, for each candidate interaction, we searched for the minimum combined *p*-value through the greedy forward inclusion of individual miRNAs. Additional miRNAs were included as candidate mediators only if they improve the joint *p*-value, as estimated using Fisher's method. MiRNAs failing to improve the joint *p*-value lack functional evidence for mediating the interaction and were thus excluded from the analysis.

### Inferred ceRNETs

We used Hermes to construct ceRNETs using matched TCGA gene and miRNA expression profiles of breast (BRCA) and prostate (PRAD) adenocarcinomas. For PRAD, this included data from 140 samples, representing 23,614 genes and 367 miRNAs[28]; while for BRCA, we used 207 samples, representing 18,748 genes and 524 miRNAs[29]. Predicted ceRNA networks included 476,456 and 447,011 interactions, for the PRAD and BRCA ceRNETs, respectively; see Tables S1-2, where each ceRNA interaction is defined by two RNAs and the miRNAs that couple them.

Due to their size, experimental validation of reverse-engineered networks is often challenging. Consequently, validation is generally performed only on a handful of interactions [30, 31] or on small subnetworks [32, 33]. To validate our inferred interactions on a more realistic scale, we used a large collection of shRNA-mediated silencing assays in the Library of Integrated Network-based Cellular Signatures (LINCS) database [14].

### The LINCS database

The LINCS database includes Luminex-based multiplexed assays to measure the expression of 1,171 genes (L1000) in response to a variety of perturbations. Selected perturbations include shRNA-mediated silencing (in triplicate) of 1,845 genes participating in ceRNA interactions, in both BRCA (MCF7) and PRAD (PC3) cell lines (Tables S3 and S4). Gene expression was measured using the L1000 assay at two time points (96h and 144h), following each perturbation. To achieve adequate statistical power, we limited our tests to genes with six or more silenced Hermes-inferred ceRNA regulators. In total, we evaluated predicted ceRNA targeting of 405 genes (of which 365 were validated at 96h and 398 at 144h) in MCF7 cells and 419 genes (of which 363 were validated at 96h and 376 at 144h) in PC3 cells.

In LINCS perturbation assays, while some data points are missing due to quality control metrics, the expression of most genes was profiled in triplicates at both 96h and 144h after shRNA transduction. On average, 3.3 unique shRNA hairpins were used to silence each of 1,845 breast and prostate oncogenes and tumor suppressors (*cancer genes*) in our networks. By definition, for each interaction, the ceRNA regulator is the one targeted by the shRNA perturbation and the ceRNA target is the one profiled after silencing to determine any interaction mediated change. Fold-change for each target, in response to the silencing of one of its regulators, was estimated by averaging across all shRNA hairpins targeting this regulator. The identities of the specific shRNA hairpins, regulators, and targets are provided in Table S5.

In total, 9,055 and 9,800 predicted BRCA interactions were tested in MCF7 cells at 96h and 144h after shRNA-mediated silencing, respectively. Similarly, 8,858 and 10,213 predicted PRAD interactions were tested in PC3 cells at 96h and 144h after silencing, respectively. Due to the small number of replicates, it is not possible to evaluate the statistical significance of individual predicted interactions. Instead, we evaluated the average effects of all ceRNA regulators in the list of silenced genes on the expression of a given ceRNA target. Average fold changes and associated standard errors were computed by comparing to non-targeting controls, at each time point, and in each relevant cellular context. Tables S3 and S4 provide mRNA expression fold change measurements following shRNA-mediated perturbation of breast and prostate cancer genes in MCF7 and PC3, respectively, based on the LINCS perturbation assays.

The size of this dataset allows for rigorous controls that avoid bias due to gene expression variability and off-target effects of RNAi-mediated perturbations. To estimate the statistical significance of target responses to gene perturbations, we examined fold change in target expression following silencing of its predicted regulator, as well as the effect of silencing a specific regulator across all of its inferred targets. For the latter, we associated each candidate target with a rank vector representing its percentile ranked relative fold-change at a specific time point following each gene silencing perturbation. We then ranked expression fold-changes of all profiled genes at a specific time point following silencing of any given gene and compared ranks across different perturbations. In this way, each candidate target was associated with a rank vector, representing its relative fold-change following each perturbation. Considering each target in isolation, we compared its response to perturbations of its predicted regulators as well as its response to shRNA-mediated silencing of other genes. Mann–Whitney U test was used to determine whether the rank of a target following silencing of its predicted regulators was lower than that following silencing of other genes.

Specifically, Fold change measurements for up to 1,171 genes in response to a given perturbation allowed ranking of the profiled genes based on the strength of the response. First, we assigned significance to the response of a gene to the silencing of its predicted regulators by comparing the set of its scores associated with perturbations of predicted regulators to the scores of all other genes, i.e. the gene's ranks following silencing of its predicted regulators vs. its ranks following silencing of all other genes. Then we used a Mann-Whitney test to determine whether the ranks of a target after silencing of its regulators was significantly lower than its ranks following silencing of all other genes. Given the number of gene perturbations (up to 1,171 gene silencing experiments), the two sets of ranks were expected to be normally distributed and can be approximated by a z-score and a corresponding p value. On average, for each target gene, the number of perturbations targeting its regulators was 1% of the total number of perturbations tested.

## Predicted cancer genes are affected by silencing of their ceRNA regulators

Figures 1B-C describe the average response of a target at a given time point to silencing of all of its tested regulators. Standard error was computed using standard error propagation techniques, i.e., the standard error was estimated as $\sqrt{\sum_{i=1}^{N} \sigma_i^2}$, where $\sigma_i$ is the standard deviation of each individual silencing experiment and $N$ is the number of tested regulators. Figures 2 and 3 show the responses of targets described in Figure 1B-C to perturbations of both predicted regulators and non-regulators in MCF7 and PC3. For perturbations of regulators, we provided p values that describe the significance of target responses. An evaluation of all individual interactions and overall responses of each target in both networks is provided in Table S5. When comparing responses of ceRNA targets to controls, as described in Figure 1B-C, we randomly assembled control sets composed of as many non-ceRNA regulators as the number of ceRNA regulators for each target. We then calculated the average fold change and the error-propagated standard error after silencing each of non-ceRNA regulator, and estimated the significance of fold changes using a two-tailed rank-sum test. This process was repeated 1,000 times to obtain averaged fold changes, propagated standard errors, and averaged p values based on the negative controls.

Results from these assays confirmed that Hermes predictions are highly enriched in bona fide ceRNA interactions in both BRCA and PRAD and that these interactions may affect the activity of key cancer genes. For instance, 10 established driver cancer genes in BRCA (*BCL2, CCND1, CCNE2, CDC42, CDKN1B, EGR1, FOS, HMGA2, NRAS* and *RB1*) and PRAD (*BCL2, CDKN1B, EGR1, HIF1A, JUN, KIT, MAP4K4, MYC, RB1* and *STAT3*) were significantly down regulated when their Hermes-inferred ceRNAs were silenced but were unaffected by silencing of negative control genes (i.e., genes not predicted as their cognate ceRNA regulators); see Figures 1B-C, 2 and 3. In total, 69% and 62% of Hermes-inferred targets were significantly down-regulated ($p < 0.05$ by U test) following shRNA-mediated silencing of their Hermes-inferred ceRNAs in MCF7 and PC3, respectively, at least at one time point; see Figure 4 and Table S5. Fold-change and *p*-values were measured by comparing average differential expression of a gene following shRNA-mediated silencing of its inferred ceRNA regulators, compared to silencing of all other genes not predicted to be ceRNA targets of these regulators. This guaranteed the most unbiased selection of negative control assays possible. Moreover, our efforts to control for specific effects that could potentially bias this comparison—including shRNA off-target effects and outliers—reaffirmed analysis results.

## Accounting for systematic biases

To ensure that comparisons of target responses to predicted regulators and non-regulators are free of bias, we repeated the analysis presented in Figure 4 for MCF7 and PC3 after controlling several properties. These included re-analyses after (1) eliminating outlier responders, i.e. ceRNA interactions associated with the most significant ceRNA-validating responses; (2) eliminating shRNAs that can act as human miRNAs[34] and produce off-target effects; and after accounting for (3) 3' UTR length and (4) CG content, (5) RMA-normalized expression, (6) expression variability, and (7) expression correlation with the predicted target. Results are presented in Figure S1 and suggest that accounting for these potential biases had relatively little effect on the number of targets that were found to be significantly down regulated by silencing their predicted regulators. We note that the average gene expression change across all interactions in LINCS is 1.0.

When eliminating outliers we removed predicted ceRNA interactions where the inducing effects were greater than Q1-1.5*STD by percentile rank from the analysis. By discarding the strongest ceRNA-like effects, we eliminated any chance that the test may be biased by a few outlier events. To eliminate potential off-target effects caused by miRNA-like behavior, we eliminated all shRNAs whose 7-base seed subsequence (2nd to 8th position) matched miRBase human miRNA seeds. To study the effects 3' UTR length and composition, we binned all potential ceRNA regulators including predicted regulators and controls; 3' UTRs were binned by either length using 25-base offsets or by GC content in 0.001 intervals. Length and content were studied independently. When comparing ceRNA interactions to non-interactions (controls), both were taken from corresponding bins. To study the effects of expression magnitude, we averaged MCF7 and PC3 gene RMA-normalized expression across 213 MCF7 and 64 PC3 experiments deposited in Gene Expression Atlas [35], binned at 0.01 intervals. Expression variability describes median absolute deviation, binned at 0.001 intervals. Expression correlation was measured using Spearman correlation with the expression profile of the target, binned at 0.01 intervals.

### Integrative Statistical Evaluation

To evaluate the significance of all tested interactions at each time point in both cell lines, we used a one-sample Kolmogorov-Smirnov test [36]. This normality test evaluates whether z scores obtained from fold-change rank comparisons follow a standard normal distribution with $\mu = 0$ and $\sigma = 1$, thus assigning significance against the null hypothesis that z scores are selected at random. We ranked targets based on z scores and calculated the expected p value when assuming that z scores were drawn from a standard normal distribution. The result is p-value estimates for each time point for each of the two cell lines, taken in aggregate across all tested interactions. The result suggested that ceRNA interactions, even when disregarding all other regulatory modalities, are highly predictive of assay observations: $p<2e-96$ and $p<1e-123$ for MCF7 at 96H and 144H, and $p<6e-84$ and $p<3e-96$ for PC3 at 96H and 144H, respectively. We did not aggregate across time points and cell lines to avoid statistical dependence that is sure to result from using the same shRNAs in multiple assays.

In MCF7 cells, of the 365 and 398 genes that could be tested at 96h and 144h following silencing of their predicted ceRNA regulators, respectively, the vast majority (i.e., 337 and 363) were down regulated. Of these, 181 and 202 were significantly down regulated ($p < 0.05$); only 28 and 35 genes were up regulated, none significantly. Similarly, for PC3 cells, 319/363 and 336/376 were down regulated (at 96h and 144h, respectively). Of these, 170 and 174 were significantly downregulated ($p < 0.05$); only 44 and 40 were up regulated, none significantly.

In total, 342 tested ceRNA targets were significantly down regulated in at least one assay, at an average of more than 2 assays per target, while none were significantly up-regulated ($p < 0.05$). In aggregate, down-regulation of predicted ceRNA targets was highly significant ($p \leq 6.0E\text{-}84$ for each time point and cell line). This analysis constitutes the largest scale validation of a regulatory network to date and suggests that hundreds of cancer genes may be altered through competition for miRNA regulation in BRCA and PRAD. Critically, these results are not merely a reflection of intrinsic coupling of gene expression in cellular systems. Indeed, equivalent numbers of interactions selected at random from non-predicted ceRNAs produced no statistically significant trends. In total, nearly 50% (50% in MCF7 and 47% in PC3) of all predicted targets were significantly down regulated by shRNA-mediated silencing of their predicted ceRNA. We note that when considering individual interactions, 31% of targets were down regulated following silencing of their predicted regulators.

## Comparisons with other ceRNA prediction methods

To test whether Hermes predictions are uniquely enriched for down regulation in LINCS data, we used LINCS assays to test predictions by MuTaME [23] and cefinder [37]. Our results suggest that while Hermes significantly outperformed both methods (Figure S2), MuTaME and cefinder predictions are significantly enriched in down regulated genes following shRNA-mediated silencing of their regulators.

We used all available predictions by cefinder and MuTaME for our comparison. cefinder scores ceRNA interactions based on the number of miRNA binding sites from the common miRNA program between the ceRNA target X and ceRNA regulator Y; only the top 50 ceRNA regulators are predicted for each ceRNA target, and Y->X doesn't imply X->Y because X might not be in the top 50 genes of Y. MuTaME provided 136 PTEN-regulating ceRNAs (of which 135 were targeted by LINCS) and the standalone program is not downloadable. Ala et al. used MuTaME to predict DICER1-regulating ceRNAs[3], but only 4 genes that were predicted to interact with DICER1 were targeted in LINCS. Consequently, we chose to compare the three methods when predicting PTEN regulators and targets (predictions by MuTaME are bidirectional), but genome-wide comparisons were made between Hermes and cefinder only. The comparisons suggest that Hermes outperforms MuTaME and cefinder when predicting PTEN targets and regulators (Figure S2), and it significantly outperforms cefinder on genome wide tests (Figure S3).

## Discussion

We proposed and re-implemented Hermes, a highly-selective context-specific method for predicting ceRNA interactions[5]. When analyzing TCGA profiles of prostate [38] and breast [39] adenocarcinomas Hermes inferred nearly 500K ceRNA interactions in each of the two tumor types.

In total, Hermes produced expression-based evidence for the regulation of over 5,000 genes by ceRNA interactions. Conclusions from perturbation assays that tested hundreds of these genes as potential targets are that half of them are dysregulated by targeting their Hermes-inferred ceRNA regulators. Put together, the results suggest that thousands of genes can be dysregulated be ceRNA interactions in each of the two contexts through exogenous perturbations or through genomic alterations that target their ceRNA regulators.

## Conclusions

Computational evidence in conjunction with high-throughput biochemical assays, suggest that ceRNA regulation is the norm and not an exception in cancer cells. While ceRNA interactions can be easily detected and validated in extreme cases—as in MYCN-amplified neuroblastomas [6] or binding-site rich RNAs [26]—they affect the expression of thousands of genes and have the potential to synergistically dysregulate drivers of tumorigenesis in multiple tumor contexts.

## Declarations

**Ethics approval and consent to participate**. Not applicable.

**Consent for publication**. Not applicable

**Availability of data and materials**. Computational methods and analyses results are fully described here. Hermes2.0 can be downloaded free of charge from its Columbia University site http://califano.c2b2.columbia.edu/hermes. LINCS assays are included in supplementary tables and can be downloaded from the LINCS website at http://www.lincsproject.org/LINCS/data.

**Competing interests**. The authors declare that they have no competing interests.

**Funding**. We acknowledge the generous funding provided by the NIH under the following grant awards: (1) Roadmap grant for a Center for the Multiscale Analysis of Genetic Networks (MAGNet) (U54CA121852), (2) Genetic Network Inference with Combinational Phenotypes (R01CA109755), (3) In Silico Research Centers of Excellence NCI-caBIG 29XS192 and 12ST1103, and (4) LINCS grants 1U01HL111566-01 and 5U01CA164184-02.

**Authors' contributions.** HSC, PS and AC designed the study. AS and TRG designed and performed assays and contributed to their analysis. HSC, MRM, MB, XY and PS analyzed data. PS and AC wrote the manuscript.

**Acknowledgements.** Not applicable

## Figures

**Figure 1.** *Model and validation of miRNA-target coupling*. (**A**) RNAs up and down regulate one another by titrating shared miRNA regulators. Up regulation of RNA B sequesters shared miRNAs, leading to weaker miRNA-mediated repression of RNA A transcripts. (**B**) In order to validate predicted interaction networks on a large scale, we evaluated whether interactions are predictive of global mRNA expression changes following shRNA perturbations using *LINCS*. A selection of known cancer genes in breast cancer and (**C**) prostate adenocarcinomas were effectively repressed following silencing of their predicted ceRNA regulators in MCF7 and PC3, respectively. Red bars represent average fold changes of a target ceRNA relative to non-targeting controls (gray bars) following silencing of its predicted ceRNA regulators at select time points; see Figures 2-3 for details. Data are represented as mean ± SEM.

Figure 2. **Target response to perturbations of both predicted ceRNA regulators and non-regulators in MCF7**. For each ceRNA target described in Figure 1B, we plot responses to shRNA-mediated silencing of (**A**) predicted ceRNA regulators and (**B**) genes not predicted to regulate each ceRNA target in MCF7. Each plot gives the profiling time point after shRNA transfection, and the total number of shRNA targets considered. For silencing of regulators, we provide p values that describe the significance of target responses shown in panel A relative to the response to silencing of other genes shown in panel B. Also provided, adjunct to each scatter plot, are box plots that describe the mean, median, 25 and 75 percentile of the distributions of ranks of the responses of this target relative to all profiled responses to shRNA perturbations.

Figure 3. **Target response to perturbations of both predicted ceRNA regulators and non-regulators in PC3**. Analogous to Figure 2, for each ceRNA target described in Figure 1C, we plot responses to shRNA-mediated silencing of (**A**) predicted ceRNA regulators and (**B**) genes not predicted to regulate each ceRNA target in PRAD.

**Figure 4. Statistical evaluation.** We plot p-values and average fold changes of target ceRNA expression following silencing of their predicted regulators, compared to silencing of all other genes in both BRCA and PRAD ceRNETs, at two profiling time

points in (**A**) MCF7 and (**B**) PC3 cells. Results for targets with six or more perturbed ceRNA regulators are shown. To estimate p values for each ceRNA target, we collected all tested regulators and compared average fold-change responses following silencing of inferred ceRNA regulators ($FC_{pos}$) vs. silencing of all other genes ($FC_{neg}$) in the network; see Figures 1-3 for illustrative example cancer genes. In total, 91% and 92% (50% and 47% significantly, at $p < 0.05$ by U test) of ceRNA targets, predicted in breast and prostate cancer, were down regulated in response to ceRNA regulator silencing in MCF7 and PC3, respectively. In total, 342 tested ceRNA targets were significantly down-regulated and none were significantly up-regulated. Comparing the number of targets with significantly low $FC_{pos}$ and $FC_{neg}$ fold changes by Mann-Whitney U-test suggests an FDR < 0.01 for overall network validation.

# Supplementary Figures

**Figures S1. Accounting for systematic biases**. To ensure that comparisons of target responses to predicted regulators and non-regulators are free of bias, we repeated the analysis presented in Figure 4 after controlling for several properties. The tables demonstrate that all results, before and after controlling for the following variables were in agreement: (1) ceRNA interactions associated with the most significant ceRNA-validating responses; (2) shRNAs that can act as human miRNAs and produce off-target effects; (3) 3' UTR length; (4) 3' UTR CG content; (5) RMA-normalized expression; (6) expression variability; and (7) expression correlation with the predicted target. Here we present comparisons to results presented in Figure 1. The number of predicted targets that were significantly ($p<0.05$) down regulated in response to transfections of shRNAs designed to target regulators are in red; down regulated ($p>0.05$) in orange; up regulated ($p>0.05$) in blue, and significantly up regulated ($p<0.05$) in green. P values give the confidence that the resulting distribution is not due to chance.

**Figure S2. The effect of predicted PTEN ceRNA regulators by each of the three methods.** Average PTEN mRNA fold change following shRNA-mediated silencing of its predicted regulators, as predicted by each ceRNA inference method including random assay selection, and inferences by Hermes, MuTaME, and cefinder. P values were calculated by comparing fold changes to random assay selection with PTEN expression profiling, using the Student's T-test (two-tailed). Average fold changes were normalized to the random assay selection. Bars show standard errors; * stands for $p<0.05$; ** for $p<0.01$; *** for $p<0.001$.

**Figure S3. Genome wide comparison**. FC comparison between Hermes, cefinder and random assay selection. Both Hermes and cefinder significantly outperform Random. Hermes outperforms cefinder at $P < 5E-07$ and $P < 2E-44$, for MCF7 and PC3, respectively; p-values based on two-sample Kolmogorov–Smirnov tests of ceRNA-target fold changes.

## Supplementary Tables

Tables S1-S2. **Inferred ceRNA networks**. Predicted interactions in the two tumor contexts: **Table S1** for BRCA, and **Table S2** for PRAD. Each table describes the coupled ceRNA pair, confidence level in the interaction, number of mediating miRNAs, and the identity of miRNA mediators.

Tables S3-4. **MCF7 and PC3 LINCS fold changes, describing gene-expression responses to perturbagens**. For each LINCS gene profiled at a given time point after each perturbation, we provide the number of replicates (Luminex plates), observed fold change relative to control, and the standard error across plates.

Table S5. **BRCA and PRAD ceRNET validation using LINCS data**. For each LINCS gene profiled at a given time point, we list (1) the number of predicted regulators that were silenced, (2) the size of the control set, which includes the gene's profiling after shRNA-mediated silencing of genes that were not predicted to target it, (3) observed and expected *U*-statistics, (4) and *Z*-statistics and associated the *p*-values obtained from them. For each interaction at a given time point, we list the target fold change in response to the perturbation, and the percentile rank when comparing target fold changes to fold changes of all other profiled genes in response to the perturbation. We also provide the identity of shRNA hairpins used.

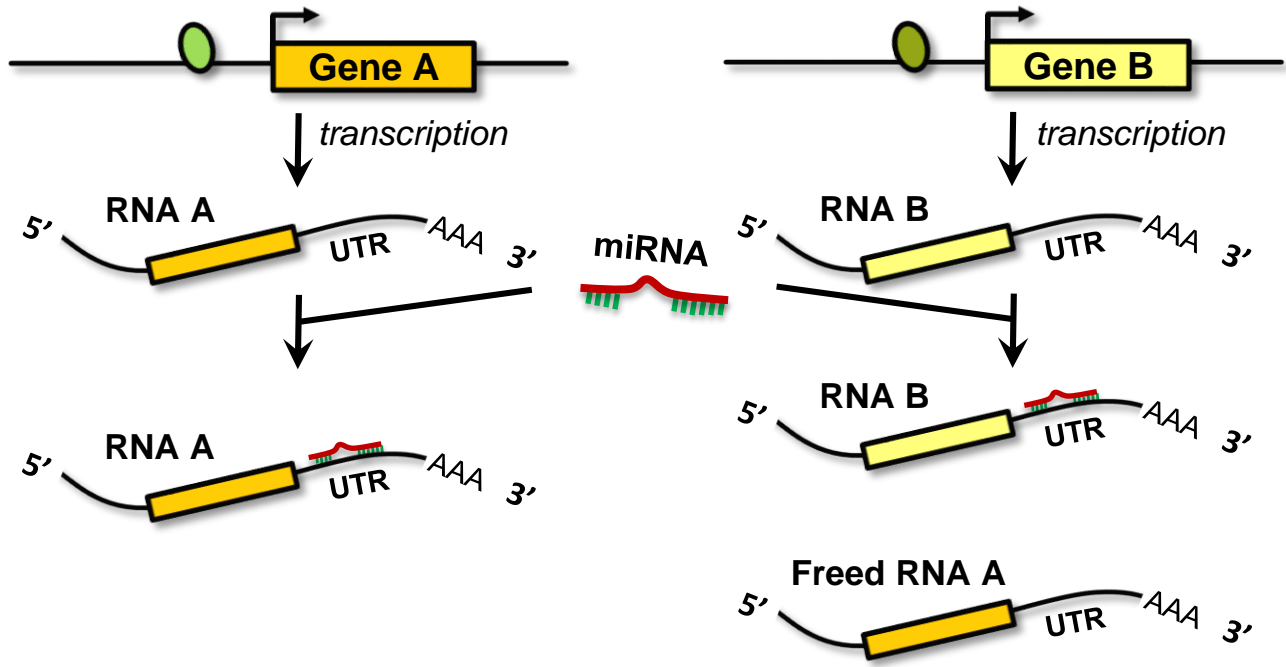
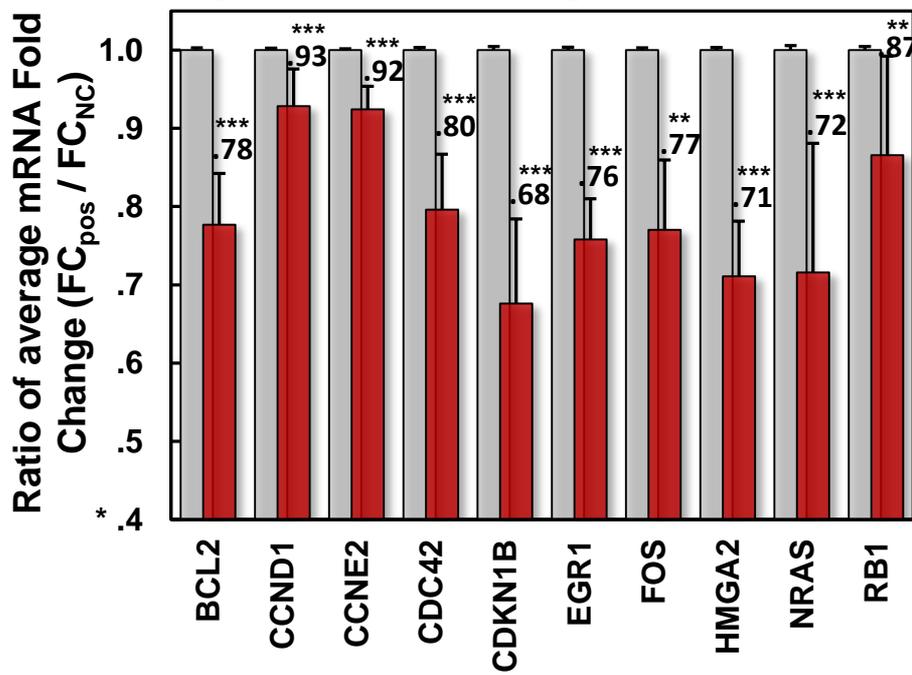
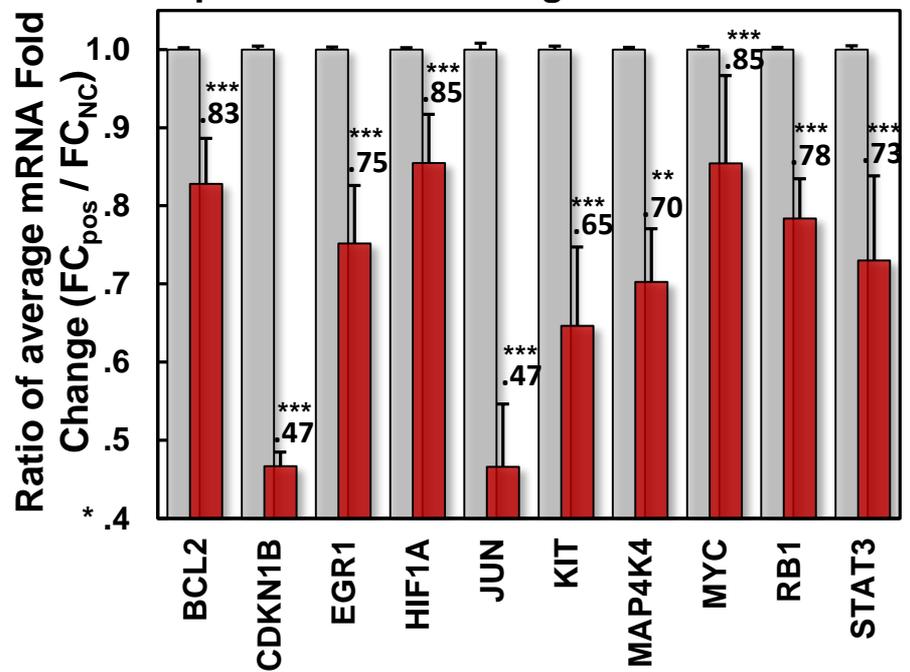

**Fig. 1**

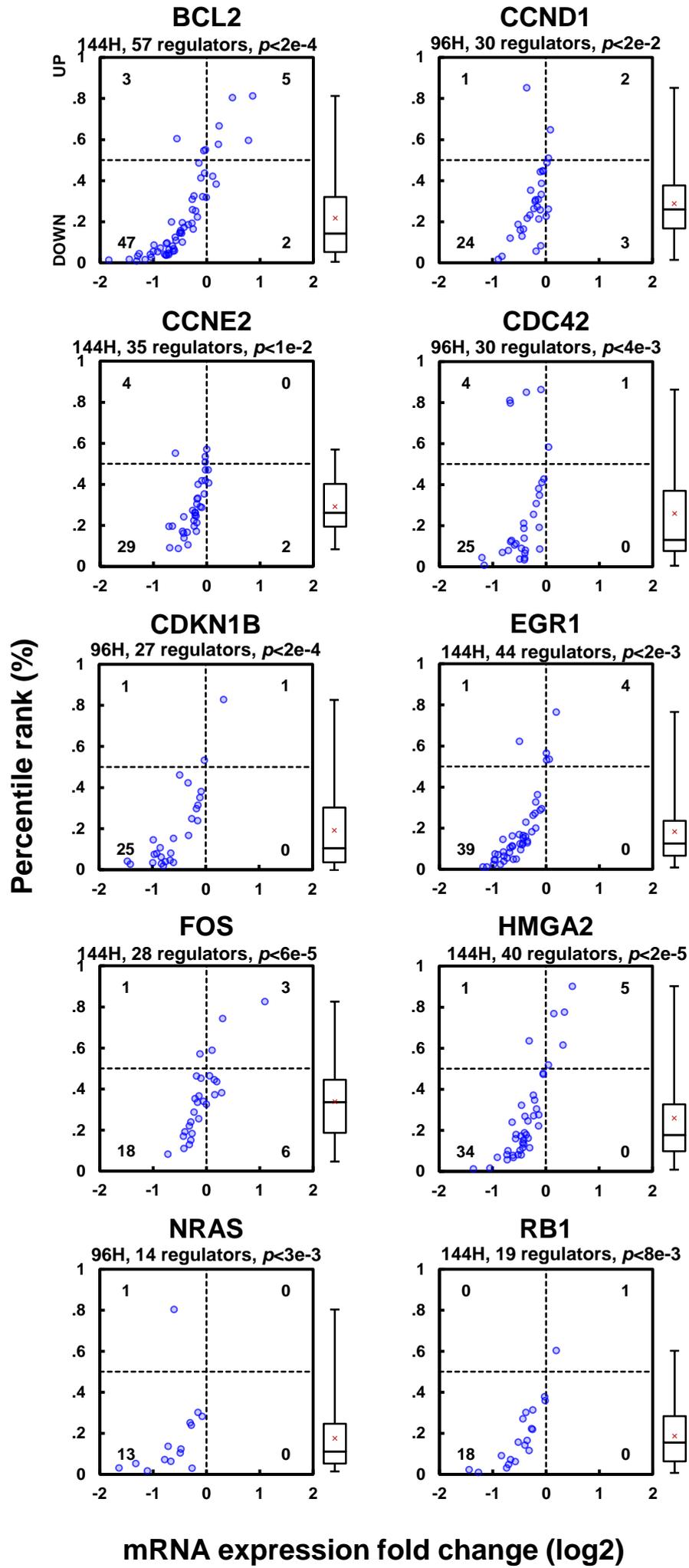 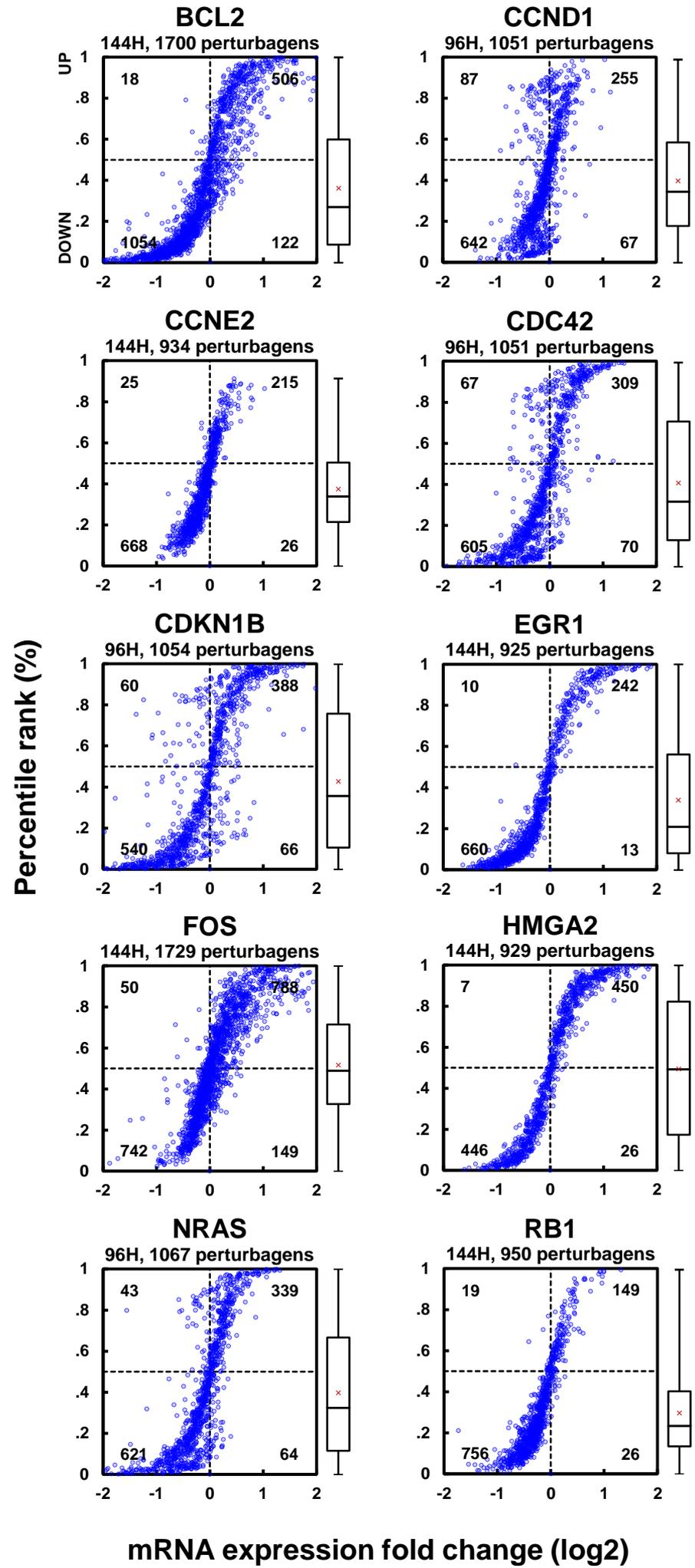

Fig. 2

**A** 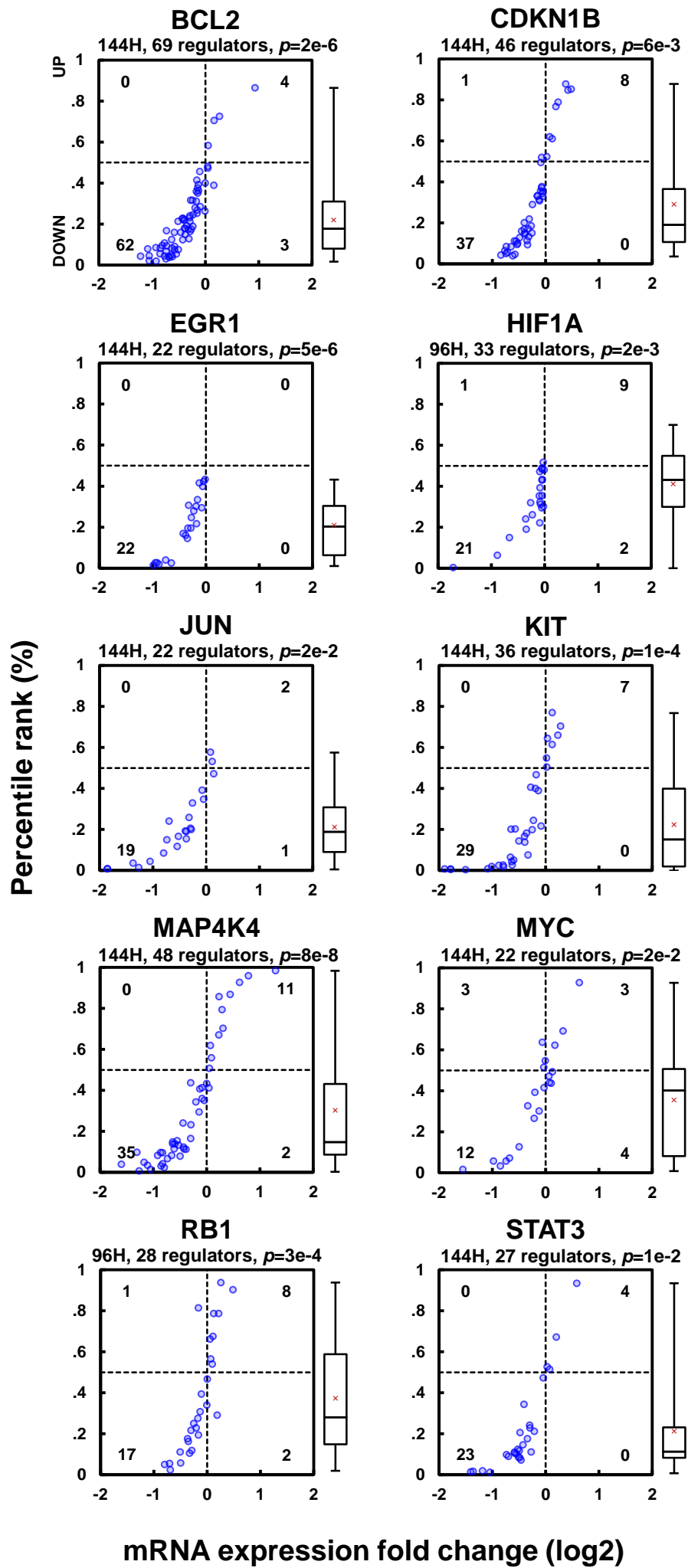 **B** 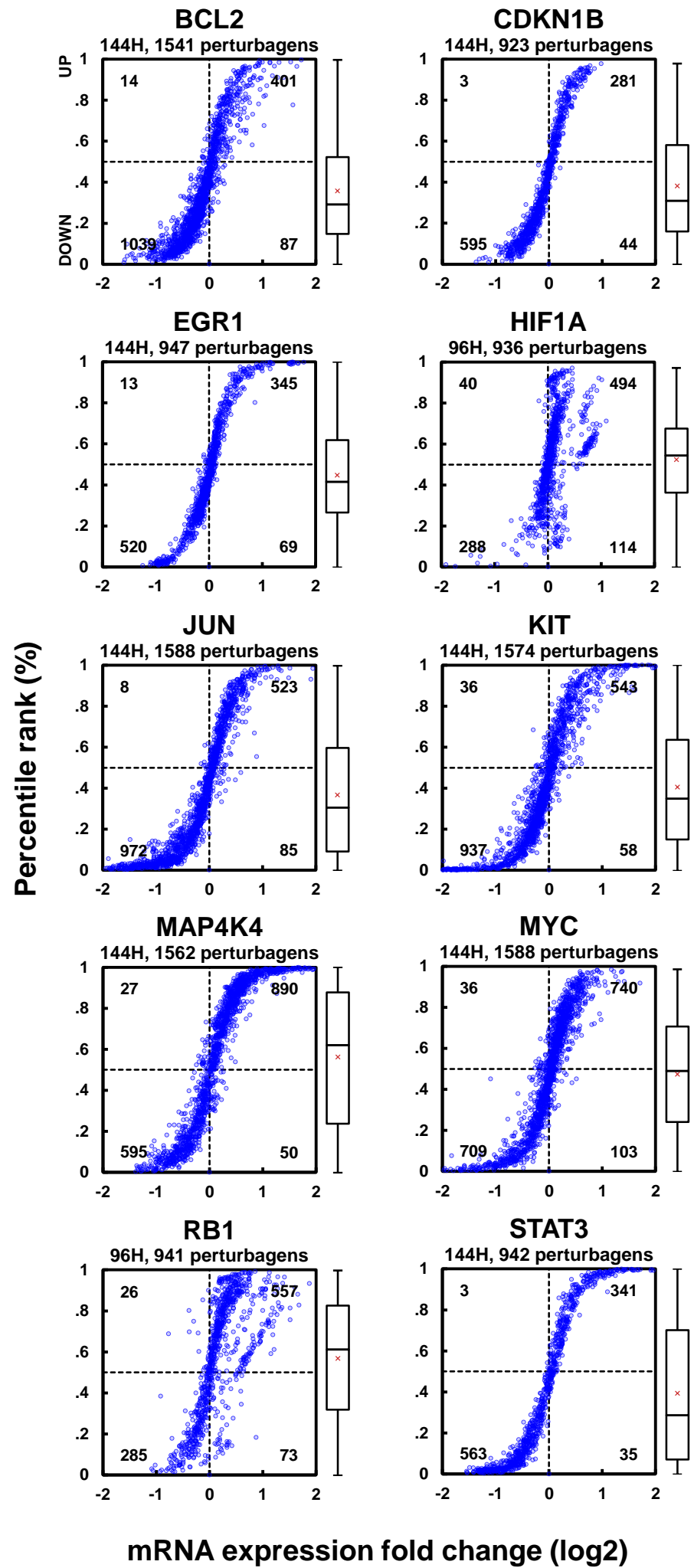

Fig. 3

**A**

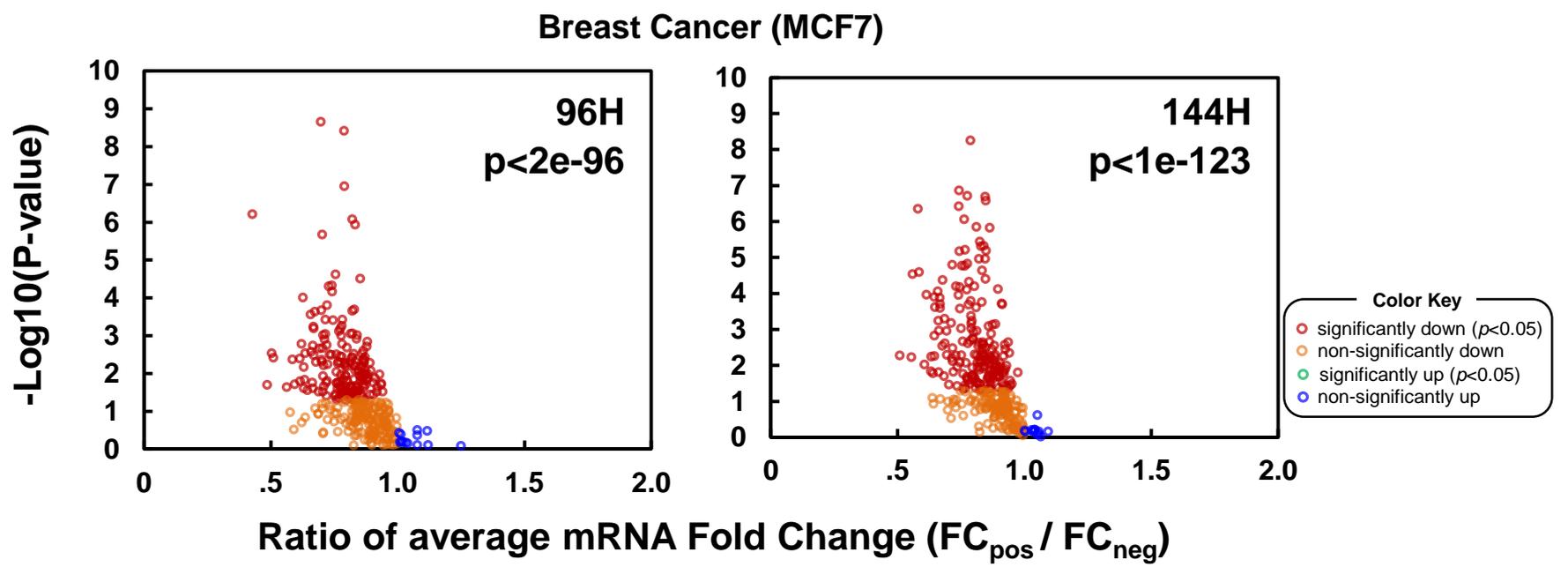

**B**

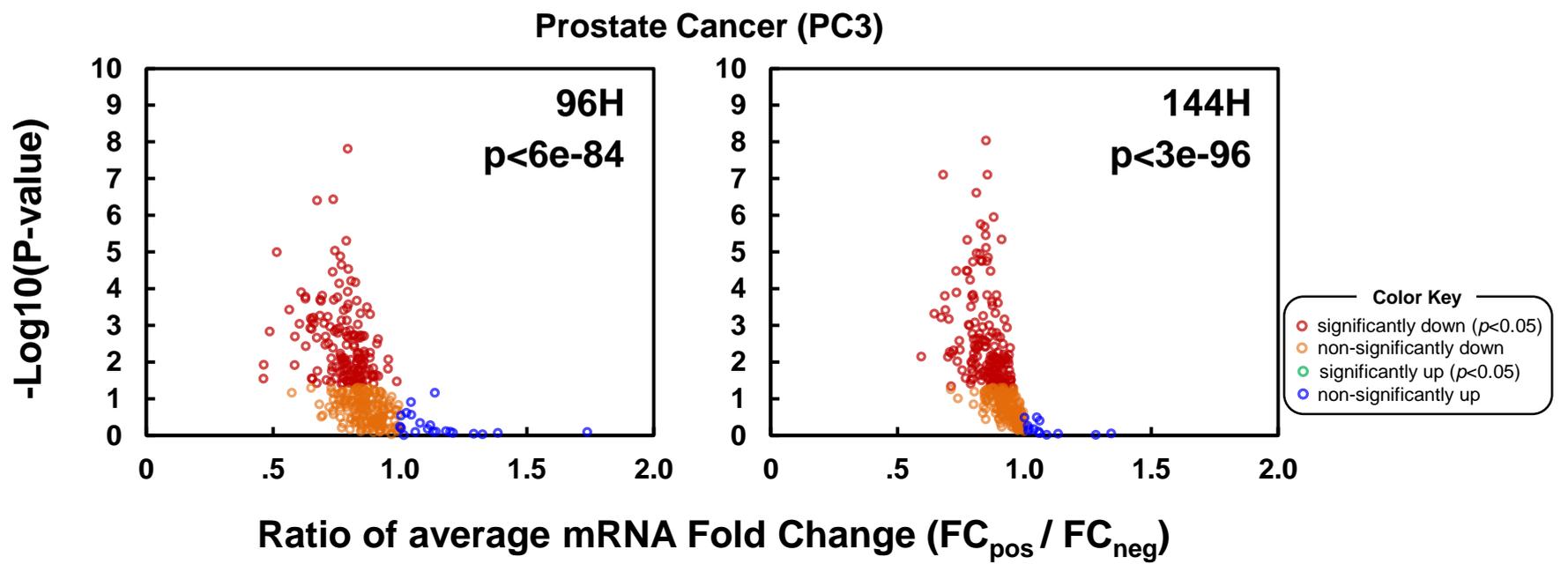

Fig. 4

| Breast (96H) | Original | Outlier free | Seed match | 3' UTR length | 3' UTR G+C content | Expression magnitude | Expression variability | Expression correlation |
|---|---|---|---|---|---|---|---|---|
| # Red | 181 | 177 | 135 | 165 | 157 | 152 | 170 | 175 |
| # Orange | 171 | 171 | 175 | 180 | 190 | 196 | 174 | 175 |
| # Blue | 13 | 16 | 15 | 19 | 18 | 16 | 20 | 14 |
| # Green | 0 | 0 | 0 | 1 | 0 | 1 | 1 | 1 |
| Total | 365 | 364 | 325 | 365 | 365 | 365 | 365 | 365 |
| P-value | 2E-96 | 9E-98 | 8E-80 | 2E-86 | 7E-89 | 8E-83 | 1E-84 | 4E-94 |

| Breast (144H) | Original | Outlier free | Seed match | 3' UTR length | 3' UTR G+C content | Expression magnitude | Expression variability | Expression correlation |
|---|---|---|---|---|---|---|---|---|
| # Red | 202 | 199 | 177 | 190 | 186 | 172 | 188 | 195 |
| # Orange | 143 | 139 | 141 | 154 | 157 | 159 | 151 | 147 |
| # Blue | 18 | 19 | 18 | 19 | 20 | 31 | 23 | 21 |
| # Green | 0 | 6 | 0 | 0 | 0 | 1 | 1 | 0 |
| Total | 363 | 363 | 336 | 363 | 363 | 363 | 363 | 363 |
| P-value | 1E-123 | 7E-121 | 5E-98 | 2E-117 | 2E-98 | 6E-101 | 6E-102 | 2E-117 |

| Prostate (96H) | Original | Outlier free | Seed match | 3' UTR length | 3' UTR G+C content | Expression magnitude | Expression variability | Expression correlation |
|---|---|---|---|---|---|---|---|---|
| # Red | 163 | 166 | 137 | 158 | 153 | 144 | 151 | 164 |
| # Orange | 174 | 172 | 161 | 167 | 176 | 177 | 172 | 163 |
| # Blue | 24 | 22 | 27 | 34 | 31 | 38 | 36 | 30 |
| # Green | 2 | 1 | 4 | 4 | 3 | 4 | 4 | 6 |
| Total | 363 | 361 | 329 | 363 | 363 | 363 | 363 | 363 |
| P-value | 6E-84 | 5E-92 | 6E-75 | 2E-72 | 5E-68 | 2E-71 | 3E-75 | 9E-80 |

| Prostate (144H) | Original | Outlier free | Seed match | 3' UTR length | 3' UTR G+C content | Expression magnitude | Expression variability | Expression correlation |
|---|---|---|---|---|---|---|---|---|
| # Red | 175 | 170 | 142 | 163 | 161 | 139 | 167 | 175 |
| # Orange | 169 | 167 | 174 | 184 | 186 | 195 | 175 | 171 |
| # Blue | 32 | 35 | 25 | 29 | 29 | 42 | 34 | 30 |
| # Green | 0 | 2 | 1 | 0 | 0 | 0 | 0 | 0 |
| Total | 376 | 374 | 342 | 376 | 376 | 376 | 376 | 376 |
| P-value | 3e-96 | 4E-84 | 6E-81 | 2E-87 | 8E-88 | 2E-90 | 6E-94 | 2E-91 |

**Fig. S1**

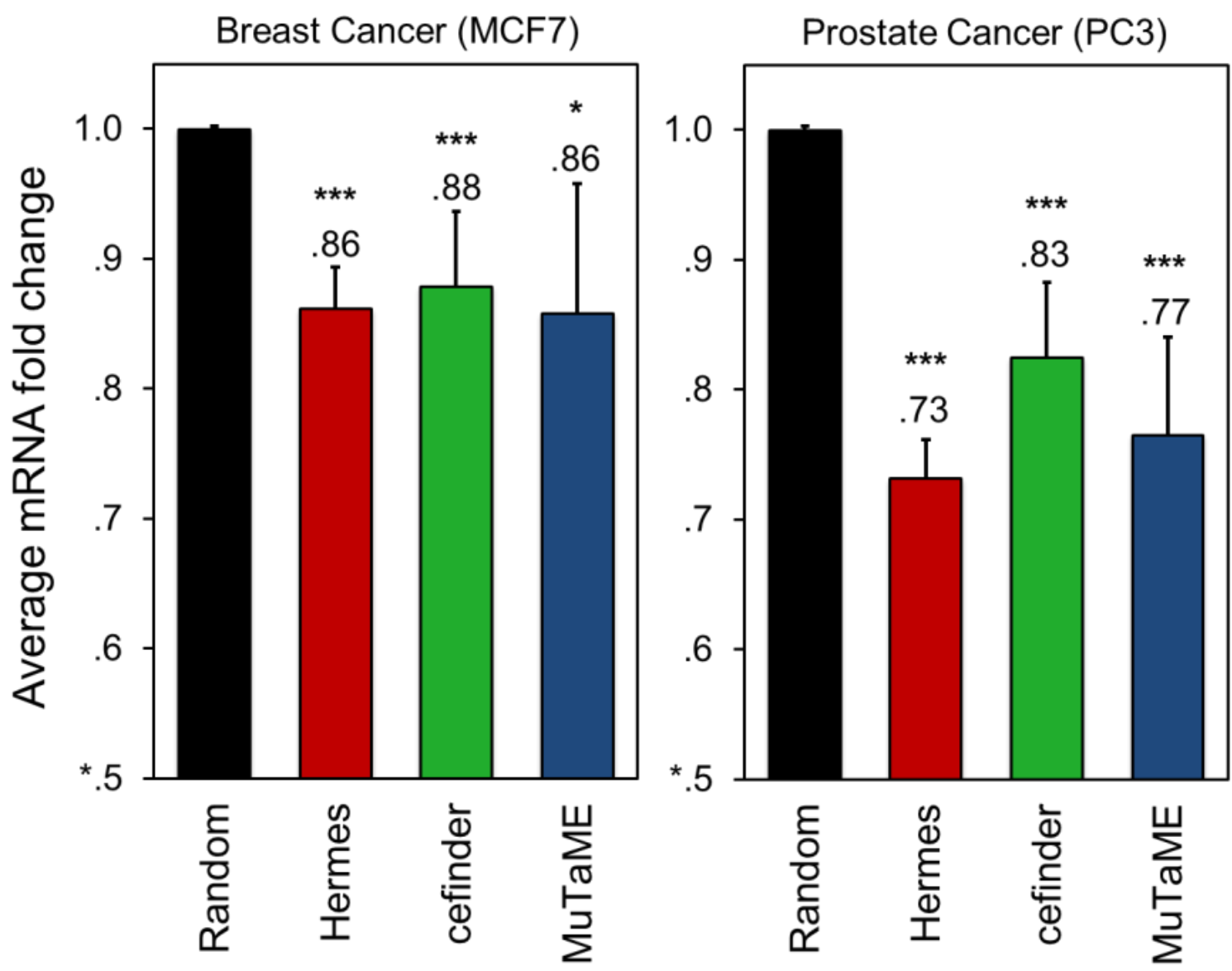

Fig. S2

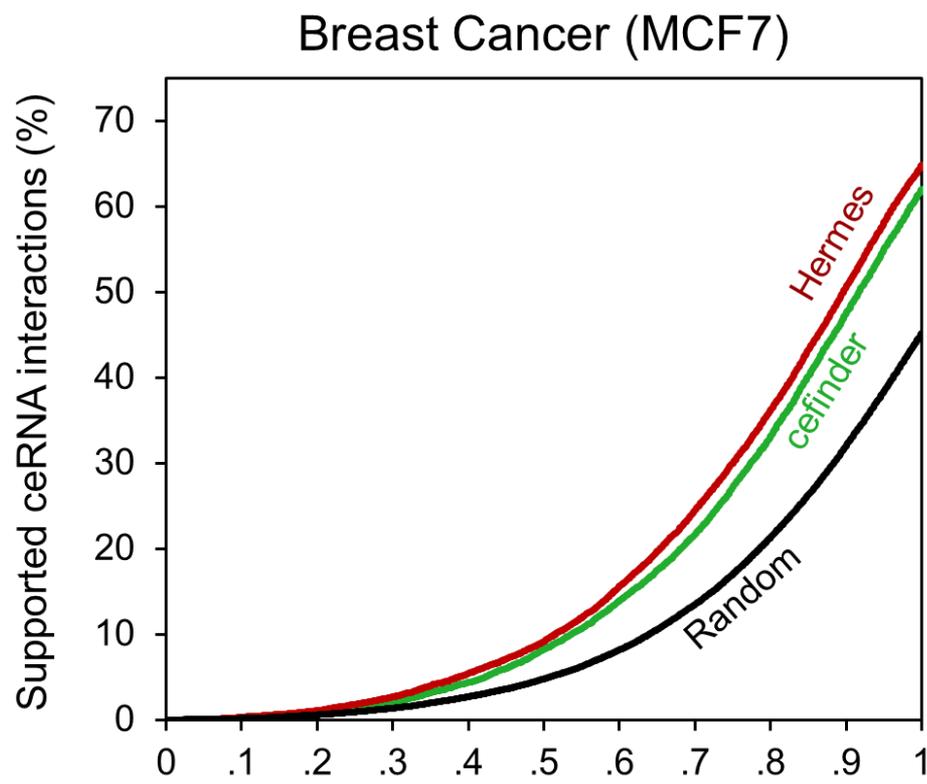

Breast Cancer (MCF7)

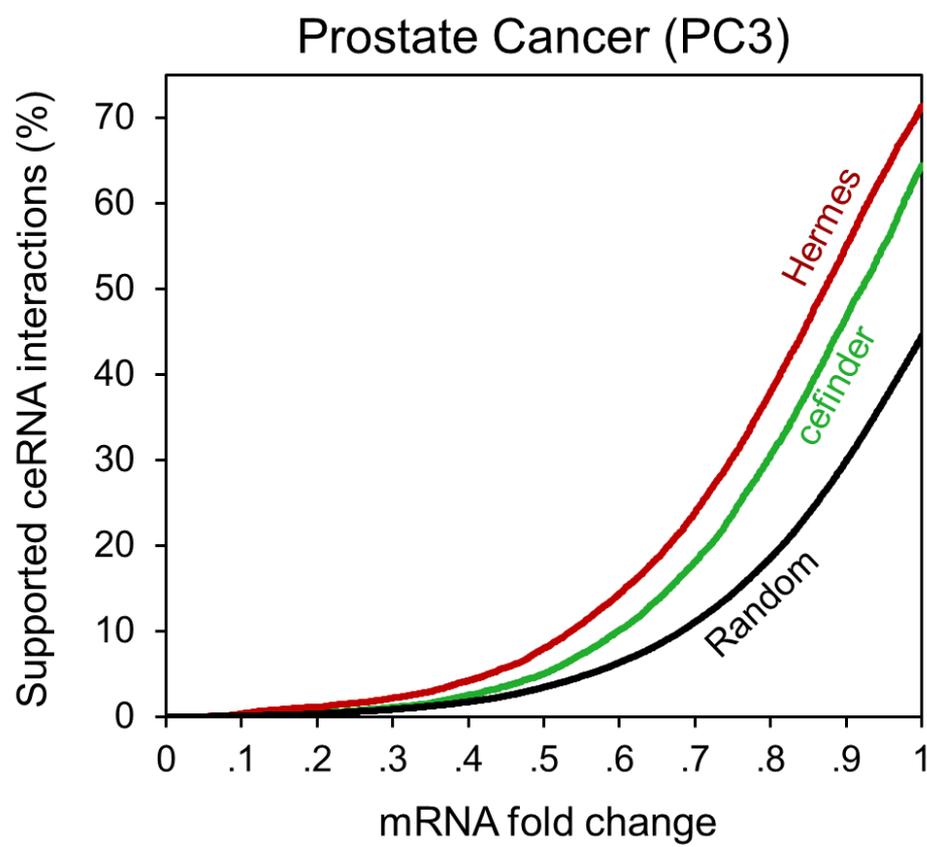

Prostate Cancer (PC3)

**Fig. S3**